# Magnetic artificial atoms based on thin-film ferrite disk particles

E.O. Kamenetskii, R. Shavit, and M. Sigalov

*Department of Electrical and Computer Engineering,
Ben Gurion University of the Negev, Beer Sheva, 84105, ISRAEL*

**Abstract**

Semiconductor quantum wells can be considered as an example of artificial atoms. Following the ideas used in the effective-mass theory, one can describe electron states in the quantum-well structure based on the Schrödinger-like equation for the envelope function. In recent years, there has been a renewed interest in high frequency dynamic properties of finite size *magnetic structures*. In a series of new publications, confinement phenomena of high-frequency magnetization dynamics in magnetic particles have been the subject of much experimental and theoretical attention. Till now, however, there are no phenomenological models of a ferrite particle with high-frequency magnetization dynamics that use the effective-mass approximation and the Schrödinger-like equation to analyze *energy eigenstates* of a whole ferrite-particle system, similarly to semiconductor quantum wells.

Magnetostatic (MS) oscillations in ferrite samples have the wavelength much smaller than the electromagnetic wavelength at the same frequency and, at the same time, much larger than the exchange-interaction spin wavelength. This intermediate position between the "pure" electromagnetic and spin-wave (exchange-interaction) processes reveals very special behaviors of the geometrical effects. The confined effects for MS oscillations in normally magnetized thin-film ferrite disks demonstrate very unique properties of artificial atomic structures.

# 1. INTRODUCTION

The multiresonance oscillations in small ferrite spheres excited by external rf magnetic fields, were experimentally observed for the first time by White and Solt in 1956 [1]. Afterwards, experiments with other-form ferrite specimens were carried out. Without an intention to survey such experiments, we would like mainly to call the reader's attention to the fundamental difference between the experimental absorption spectrums of the sphere-form [1] and the disk-form ferrite resonators [2,3]. The $\delta$ - functional character of the multi-resonance spectra, one can see in a case of a ferrite disk resonator, leads to a clear conclusion that the energy of a source of a DC bias magnetic field is absorbing "by portions" or discretely, in other words. Contrary, the spectrum of a ferrite sphere is characterized by a very few and very "spreading" absorption peaks. What can be the nature of such a strong discreteness of the ferrite-disk-resonator spectrum? Certainly, there should be a certain inner mechanism of the quantization of the DC energy absorbed by a small disk-form ferrite sample.

The energy states of a spin system coupled by exchange interactions are wavelike. Spin waves are intrinsic excitations in magnetic materials. The energy of a spin wave is quantized, and the unit of energy of a spin wave is called a *magnon*. On the corpuscular language, an oscillating process in a magnetically ordered body is a collection of magnons [4]. The magnetostatic (MS) - wave oscillations in ferrite samples are described with neglect of the exchange interaction [4-6]. In paper [7] we showed that in a ferrite disk resonator, MS oscillations are characterized by energy eigenstates, which are not described as the magnetization-field quantization. In this paper we will show that the process of MS-wave propagation in such resonators can be considered as the motion process of certain quasiparticles having quantization of energy and characterizing by effective masses. Conventionally, these quasiparticles are called the "*light magnons*". As we discussed in [7], in a normally magnetized ferrite disk the MS-potential wave function can be considered as the probability distribution function. The "light magnon" distribution is defined as the probability density distribution function. The confined phenomena of the "light magnon" oscillations in normally magnetized thin-film ferrite disks demonstrate very unique properties of artificial atomic structures.

## 2. ON THE MAGNETOSTATIC DESCRIPTION OF OSCILLATING MODES IN FERRITE SAMPLES

For alternative electromagnetic fields, the quasistatic processes are characterized as follows. The quasistatics means the fact that the characteristic specimen size $l$ is much less than the free-space electromagnetic wavelength $\lambda \sim c/\omega$ [8]:

$$l << c/\omega \tag{1}$$

In this case, the magnetic field distribution inside and outside a sample is described by the static-field equations:

$$\nabla \times \vec{H} = 0, \tag{2a}$$

$$\nabla \cdot \vec{B} = 0 \tag{2b}$$

In such a description, the Maxwell equation

$$\nabla \times \vec{E} = -\frac{\partial \vec{B}}{\partial t} \tag{3}$$

has only a *subsidiary purpose* just to express the magnetic field energy. Really, the magnetic field does not perform any work over moving charges. So *to calculate the energy variation one should take into account the electric field induced in accordance the Faraday law* (3). In quasistatic processes in conductive media, for example, the energy variation is considered as a work, which an electric field produces over the currents – the sources of the magnetic field [8].

Magnetostatic-wave oscillations in a ferrite sample occupy a special place between the "pure" electromagnetic and spin-wave (exchange) processes. For the characteristic specimen size $l$, MS-wave oscillations take place when

$$\sqrt{\alpha} \ll l \ll c/\omega \qquad (4)$$

where $\sqrt{\alpha}$ is the characteristic length of the exchange energy [8]. Relation (4) means that we neglect all the electromagnetic effects (the effects connected with a finite velocity of propagation of electromagnetic perturbations) and, at the same time, all the exchange processes (the effects connected with spatial dispersion in ferrites). The energy density of MS-wave oscillations is not the electromagnetic-wave density of energy and not the exchange energy density as well.

MS waves and oscillations in ferrites are described based on the Walker equation – the second-order differential equation written for MS potential $\psi$ $(\vec{H} = -\nabla \psi)$. This equation is derived from the Landau-Lifshitz motion equation and two differential equations (2a) and (2b). In an unbounded ferrite, MS waves at the oscillating frequencies have any wavelength. The frequency degeneracy is removed by considering the effects of finite sample boundaries. The boundary conditions are that the tangential component of $\vec{H}$ and the normal component of $\vec{B}$ should be continuous across the boundary of the specimen. For many applications, MS waves in ferrite samples are excited by current transducers. The product of the current in such a transducer multiplied by the curl electric field [see Eqn. (3)] provides the energy for the MS-wave excitation [5,6].

For monochromatic MS-wave processes, Eqn (3) can be written as:

$$\nabla \times \vec{E} = -i\omega \vec{B} = i\omega \, \ddot{\mu}(\omega) \cdot \nabla \psi , \qquad (5)$$

where $\ddot{\mu}$ is the permeability tensor. Let us formally consider the right-hand side of Eqn. (5) as the *magnetic displacement current*. We can write

$$\nabla \times \vec{E} = -\vec{j}_{dis}^{\,m}, \tag{6}$$

where

$$\vec{j}_{dis}^{\,m} \equiv i\omega\vec{B}. \tag{7}$$

The physical meaning of the magnetic displacement current arises from the MS nature of the observed oscillations. In these oscillations, the time-varying electric field cannot be a "source" of the magnetic field (because of neglect of the electric displacement current). On the contrary, the existing magnetic displacement current is a "source" of the electric field. Since $\nabla \cdot \vec{B} = 0$, one has

$$\nabla \cdot \vec{j}_{dis}^{\,m} = 0. \tag{8}$$

Defining the curl electric field in terms of a magnetic vector potential:

$$\vec{E} = -\nabla \times \vec{A}^m, \tag{9}$$

one obtains:

$$\nabla^2 \vec{A}^m = -\vec{j}_{dis}^{\,m} = i\omega\,\vec{\mu}(\omega) \cdot \nabla\psi \tag{10}$$

This is the Poisson equation for magnetic vector potential $\vec{A}^m$ in the MS description.

For the potential magnetic and curl electric fields, the energy balance equation for monochromatic MS waves is written as

$$\nabla \cdot \left(\vec{E} \times \vec{H}^* + \vec{E}^* \times \vec{H}\right) = \vec{H} \cdot \left(i\omega\vec{B}^*\right) - \vec{H}^* \cdot \left(i\omega\vec{B}\right) = \vec{H} \cdot \left(\vec{j}_{dis}^{\,m}\right)^* - \vec{H}^* \cdot \vec{j}_{dis}^{\,m} \tag{11}$$

This expression shows, in particular, that in a ferrite sample MS waves can be excited by external rf magnetic field $\vec{H}$. The excitation integral can be defined as $\int_V \vec{H} \cdot \vec{B}^* dV$ [9]. This differs from the excitation integral $\int_V \vec{H} \cdot \vec{m}^* dV$ used in [6]. So the magnetization field does not define the energy balance for MS waves $\vec{m}$. The energy states of MS waves should be analyzed based on the magnetic flux density field $\vec{B}$.

MS oscillations in a ferrite sample can be completely described by the complex-value scalar-potential wave functions. In a case of a normally magnetized ferrite disk one has a complete-set discrete spectrum of MS-potential eigen wave functions, which can be considered as the probability distribution functions. The power flow density along axial axis of a disk (expressed by the potential magnetic and curl electric fields) is the flow density of probability [7]. These features cause the main properties of the "light-magnon" oscillations in an axially magnetized ferrite disk.

## 3. MAGNONS AND MS OSCILLATIONS

The energy of a spin wave is quantized, and the unit of energy of a spin wave is called a *magnon*. On the corpuscular language, a spin-wave oscillating process in a magnetically ordered body is a collection of magnons. The magnons, being the quantum quasiparticles, are characterized as the quantized states with the lack of localization in space. For frequency of magnetic oscillations $\omega$, the energy of every magnon is:

$$\varepsilon = \hbar\omega, \qquad (12)$$

Suppose that the characteristic size $l$ of a ferrite specimen is determined by conditions (4). In this case, for short-wavelength (spin-wave) magnetic oscillations the role of boundary conditions is negligibly small. Momentum is defined as

$$\vec{p} = \hbar\vec{k}, \qquad (13)$$

where $\vec{k}$ is a wave vector of a spin wave. The correlation between energy and a momentum of a magnon is similar to such a correlation for a free non-relativistic particle [4-6].

For long-wavelength (the MS, or Walker-type) magnetic oscillations one does not have so simple connection between the magnon energy and momentum since the wavenumbers are strongly dependent on the ferrite sample geometry. In the MS description, the rf magnetization is defined as

$$\vec{m} = -\vec{\tilde{\chi}}(\omega) \cdot \nabla \psi, \tag{14}$$

where $\vec{\tilde{\chi}}$ is the magnetic susceptibility tensor. Considering the magnon number in the whole of a ferrite sample, one can define the summary energy of magnons in $q$-th MS mode as:

$$W_q = n_q \varepsilon_q, \tag{15}$$

where $n_q$ is the magnon number in $q$-th MS mode. The magnon energy is defined as

$$\varepsilon_q = \hbar \omega_q, \tag{16}$$

where $\omega_q$ is the frequency of mode $q$. For modes with a circular precession, the number of magnons in $q$-th MS mode can be estimated as:

$$n_q = \frac{1}{2 M_0 \gamma \hbar} \int_V m_q^2 \, dV \tag{17}$$

where $m_q^2$ is a square of the alternative magnetization amplitude, $\gamma$ is the gyromagnetic ratio, and $M_0$ is the saturation magnetization. This formula was obtained as a result of equality between the summary energy of magnons and the oscillating energy of the rf magnetization in a ferrite sample [6]. Is it possible to quantize the magnetization field in MS oscillations? For a ferromagnetic spheroid with the dc magnetic field directed along $z$ axis, Walker obtained the orthogonality relation for two oscillating MS modes [10]:

$$[\omega^{(\lambda)} + (\omega^{(\nu)})^*] \int_V [\vec{m}_\perp^{(\lambda)} \times (\vec{m}_\perp^{(\nu)})^*] \cdot \vec{e}_z \, dV = 0 \tag{18}$$

Considering this relation, Walker discussed the problem if the MS modes actually diagonalize the total magnetic energy of a sample [10].

Concerning the above question about possible quantization of magnetization for MS oscillations, we have to express, however, a certain criticism. This is the fact that in a case of MS oscillations one cannot formulate any spectral problem for the magnetization field. The energy eigenvalue problem for MS oscillations can be formulated based on not the magnetization field, but the MS-wave potential function.

## 4. ENERGY EIGENVALUE PROBLEM FOR MS OSCILLATIONS IN A FERRITE DISK RESONATOR

We have discussed above that taking into account the well-known fact that the energy of a spin wave is quantized, and that a magnon is a quasiparticle characterizing the unit of the spin-wave energy, one cannot formulate the energy eigenvalue problem for MS oscillations based on the magnetization field. Now the question arises: By what manner is it possible to quantize the MS-wave energy?

In a normally magnetized ferrite-disk resonator with a small thickness to diameter ratio, the monochromatic MS-wave potential function $\psi$ is represented as [7]:

$$\psi = \sum_{p,q} A_{pq} \widetilde{\xi}_{pq}(z) \widetilde{\varphi}_q(\rho,\alpha), \qquad (19)$$

where $A_{pq}$ is a MS mode amplitude, $\widetilde{\xi}_{pq}(z)$ and $\widetilde{\varphi}_q(\rho,\alpha)$ are dimensionless functions describing, respectively, "thickness" ($z$ coordinate) and "in-plane", or "flat" (radial $\rho$ and azimuth $\alpha$ coordinates) MS modes. For a certain-type "thickness mode" (in other words, for a given quantity $p$), every "flat mode" is characterized by its own function $\widetilde{\xi}_q(z)$.

Because of separation of variables, one can impose independently the electrodynamical boundary conditions – the continuity conditions for the MS potential $\psi$ and for the normal component of the magnetic flux density – on a lateral cylindrical surface ($\rho = R$, $0 \leq z \leq h$) and plane surfaces ($z = 0$, $z = h$). To analyze eigenstates of MS oscillations in a ferrite disk, we should start with consideration of the states of MS waves in an axially magnetized ferrite rod. After that we can extend our analysis for a normally magnetized ferrite disk.

Let us consider an axially magnetized ferrite rod. MS oscillations in this one-dimensional linear structure are described by scalar wave function $\psi$. It is known that, by appropriate change of variables, any system of equations describing oscillations in *one-dimensional linear structures with distributed parameters* may be written as (see, for example, [11]):

$$\hat{Q}\vec{u} = \frac{\partial \vec{u}}{\partial t}, \tag{20}$$

where $\vec{u}(z,t)$ is a vector function with components $u_1, u_2, ...$ describing the system properties and $\hat{Q} = \hat{Q}(z)$ is a differential-matrix operator. In a general case, the $jk$ matrix element of operator $\hat{Q}(z)$ is written as:

$$Q_{jk}(z) = a_{jk}^{(m)}(z)\frac{\partial^m}{\partial z^m} + a_{jk}^{(m-1)}(z)\frac{\partial^{m-1}}{\partial z^{m-1}} + ... + a_{jk}^{(1)}(z)\frac{\partial}{\partial z} + a_{jk}^{(0)}(z), \tag{21}$$

where $m$ is an order of every differential equation of a system (20). Since MS oscillations in a ferrite rod are described by scalar wave function $\psi$, for a lossless structure one has from Eqns. (20) and (21):

$$a^{(1)}(z)\frac{\partial^2 \psi(z,t)}{\partial z^2} + a^{(2)}(z)\psi(z,t) = \frac{\partial \psi(z,t)}{\partial t} \tag{22}$$

This is the Schrödinger-like equation. The solution of this equation can be found as a product of functions dependent only on $z$ and only on $t$. Since the left-hand side of Eqn. (22) is the function dependent only on $z$ and the right-hand side – only on $t$, one can conclude that both these sides should be equal to the same constant value. This makes possible to consider Eqn. (22) as the *stationary-state* equation.

Let a one-dimensional linear structure be a waveguide structure with parameters not dependent on longitudinal $z$ coordinate. So coefficients $a^{(1)}$ and $a^{(2)}$ in Eqn. (22) are not dependent on $z$. For harmonic processes, coefficients $a^{(1)}$ and $a^{(2)}$ should be imaginary quantities. Based on the energy balance equation [7], one obtains the average energy of MS mode $n$ in a waveguide section restricted by coordinates $z_1$ and $z_2$:

$$\overline{W}_n = -\frac{1}{4a_n^{(1)}} i\omega\mu_0 \int_{z_1}^{z_2}\int_S \psi_n \psi_n^* \, ds \, dz + C \tag{23}$$

where $C$ is an arbitrary quantity not dependent on time. We can normalize the process in a supposition that constant $C$ is equal to zero. One can see that coefficient $a_n^{(2)}$ is not included in the expression of average energy. The only coefficient included in this expression is coefficient $a_n^{(1)}$. Another important conclusion following from Eqn. (23) is that for any coefficient $a_n^{(1)}$ the energy can be orthogonolized with respect to the known $\psi_n$ eigenfunctions.

Let us represent the MS-potential function in a ferrite rod as

$$\psi = A\tilde{\varphi}\, e^{-i\beta z} \tag{24}$$

where $A$ is a dimensional coefficient and $\tilde{\varphi}$ is a dimensionless membrane function. Since membrane functions of MS modes in an axially magnetized ferrite rod give a complete discrete set of functions (on a waveguide cross section) [7,12], the dimensionless membrane function $\tilde{\varphi}$ can be written as

$$\tilde{\varphi} = \sum_{n=1}^{\infty} b_n \tilde{\varphi}_n, \tag{25}$$

where $\tilde{\varphi}_n$ is a membrane function of MS mode and $b_n$ are constants. In a case of a cylindrical ferrite rod, $\tilde{\varphi}_n$ are described by the Bessel functions [12]. The wave function $\tilde{\varphi}$ is normalized to unity when the coefficients $b_n$ satisfy the relation $\sum_n |b_n|^2 = 1$ [7].

Based on the Walker equation, one has for every MS mode in an axially magnetized rod [7]:

$$\hat{G}_\perp \tilde{\varphi}_n = \beta_n^2 \tilde{\varphi}_n, \tag{26}$$

where

$$\hat{G}_\perp \equiv \mu \nabla_\perp^2, \qquad (27)$$

$\nabla_\perp^2$ is the two-dimensional, "in-plane", Laplace operator. For propagating MS modes, operator $\hat{G}_\perp$ is the positive definite operator.

The fact that coefficient $a^{(2)}$ is not included in the expression of average energy gives us a possibility to consider different cases based on certain physical models. One can see that when $a^{(2)} \equiv 0$, Eqn. (22) resembles the Schrödinger equation for "free particles". This is the case of a constant value of bias magnetic field $\vec{H}_0$. Certainly, when a ferrite specimen (having saturation magnetization of a ferrite material), is placed into a bias magnetic field, one has a constant "potential energy" of this ferrite sample in the DC magnetic field. For a given constant value of a bias magnetic field $\vec{H}_0$, the spectral properties of a structure are exhibited with respect to frequency $\omega$. For monochromatic MS wave and when $a_n^{(2)} \equiv 0$, coefficients $a_n^{(1)}$ are found as [see Eqn. (22)]:

$$a_n^{(1)} = -\frac{i\omega}{\beta_n^2}, \qquad (28)$$

We define a notion of the *normalized average MS energy of mode n* as the average (on the RF period) energy of MS waveguide section with unit length and unit characteristic cross section. This energy for a mode with unit amplitude $\left(|b_n|^2 = 1\right)$ is expressed based on Eqn. (23) as:

$$E_n = \frac{1}{4} g \mu_0 \beta_n^2 \qquad (29)$$

where $g$ is the unit dimensional coefficient, having the same dimension as a squared amplitude $A^2$ [see Expr. (24)].

Based on the above consideration of the states of MS waves in an axially magnetized ferrite rod, we extend now our analysis to a case of a normally magnetized ferrite disk. In a ferrite disk with a small thickness/diameter ratio, the spectrum of "thickness modes" is very "rare" compared to the "dense" spectrum of "flat modes" [7]. So, the spectral properties of such a resonator can be entirely described based on consideration of only a fundamental "thickness mode" [see Eqn. (19)]. Formally, to find an average energy of MS mode $q$ in a ferrite disk resonator, one can extend Eqn. (23) as follows:

$$\overline{W}_q = -\frac{1}{4} i \omega \mu_0 \int_S \left[ \frac{1}{\left(a_q^{(1)}\right)^{(D)}} \int_{-\infty}^{0} \psi_q \psi_q^* dz + \frac{1}{\left(a_q^{(1)}\right)^{(F)}} \int_{0}^{h} \psi_q \psi_q^* dz + \frac{1}{\left(a_q^{(1)}\right)^{(D)}} \int_{h}^{\infty} \psi_q \psi_q^* dz \right] ds + C \quad (30)$$

where superscripts *(D)* and *(F)* mean, respectively, the dielectric $(-\infty \leq z \leq 0; h \leq z \leq \infty)$ and ferrite $(0 \leq z \leq h)$ regions.

It becomes clear, however, that only the term corresponding to the ferrite region (where we have propagating MS waves) gives the real quantity. Since for a "flat" mode $q$ in a normally magnetized ferrite disk there are two waves propagating forth and back with respect to $z$-axis, the average energy will be twice more than the energy expressed by Eqn. (29). We have in result (in a supposition that constant $C$ is equal to zero):

$$E_q = \frac{1}{2} g \mu_0 \left(\beta_q^{(F)}\right)^2 \quad (31)$$

where $\beta_q^{(F)}$ is a MS-wave propagation constant along z-axis in a ferrite region $(0 \leq z \leq h)$.

Since a two-dimensional ("in-plane") differential operator $\hat{G}_\perp$ contains $\nabla_\perp^2$ (the two-dimensional, "in-plane", Laplace operator), a double integration by parts (the Green theorem) on $S$ – a square of an "in-plane" cross section of an open ferrite disk – of the integral $\int (\hat{G}_\perp \widetilde{\varphi}) \widetilde{\varphi}^* dS$, gives the following boundary condition for the energy orthonormality:

$$\mu \left(\frac{\partial \widetilde{\varphi}}{\partial \rho}\right)_{\rho=R^-} - \left(\frac{\partial \widetilde{\varphi}}{\partial \rho}\right)_{\rho=R^+} = 0 \quad (32a)$$

or

$$\mu(H_\rho)_{\rho=R^-} - (H_\rho)_{\rho=R^+} = 0, \qquad (32b)$$

where $R^-$ and $R^+$ designate, respectively, the inner (ferrite) and outer (dielectric) regions of a disk resonator with radius $R$.

For operator $\hat{G}_\perp$, the boundary condition of the MS-potential continuity together with boundary condition (32a) [or (32b)] are the so-called *essential* boundary conditions [13]. When such boundary conditions are used, the MS-potential eigen functions of operator $\hat{G}_\perp$ form a *complete basis in an energy functional space*, and the functional describing an average quantity of energy, has a minimum at the energy eigenfunctions [13]. The essential boundary conditions differ from the homogeneous electrodynamics boundary conditions at $\rho = R$, which demand continuity for $\tilde{\varphi}$ and continuity for the radial component of the magnetic flux density. The last ones are called as *natural* boundary conditions [13].

To find the normalized average MS energies of mode $q$ determined by Eqn. (31), one should find the MS-wave propagation constants $\beta_q^{(F)}$. Since in an open ferrite disk with a small thickness/diameter ratio one can use separation of variables [7], the propagation constants are defined as solutions of a system of two equations:

$$\tan(\beta^{(F)} h) = -\frac{2\sqrt{-\mu}}{1+\mu} \qquad (33)$$

and

$$(-\mu)^{\frac{1}{2}} \frac{J'_\nu}{J_\nu} + \frac{K'_\nu}{K_\nu} = 0 \qquad (34)$$

where $\mu$ is the diagonal component of the permeability tensor, $h$ is the thickness of a ferrite disk, $J_\nu, J'_\nu, K_\nu,$ and $K'_\nu$ are the values of the Bessel functions and their derivatives on a lateral cylindrical surface ( $\rho = R$, $0 \leq z \leq h$ ). Eqn. (33) corresponds to characteristic equation for MS waves in a normally magnetized ferrite slab [14]. Eqn. (34) is the characteristic equation for MS waves in an axially magnetized ferrite rod [12], but with the essential boundary conditions. With use of essential boundary conditions one obtains solutions *not dependent on a sign of* $\nu$ - the order of the Bessel function.

Now we can formulate the energy eigenvalue problem for MS waves in a ferrite disk resonator as the problem defined by the differential equation:

$$\hat{F}_\perp \tilde{\varphi}_q = E_q \tilde{\varphi}_q \qquad (35)$$

together with the corresponding (essential) boundary conditions. A two-dimensional ("in-plane") differential operator $\hat{F}_\perp$ is determined based on Eqns. (26), (27) and (31) as follows:

$$\hat{F}_\perp = \frac{1}{2} g \mu \mu_0 \nabla_\perp^2 \qquad (36)$$

## 5. ON QUANTIZATION OF MAGNETIZATION FOR MS OSCILLATIONS

We have shown that in a ferrite disk resonator every MS-wave oscillating mode is characterized by a certain energy level. Now we should come to the question: What can be a correlation between the analyzed above energies of MS oscillations and the density of magnetization energy? In other words: Could the oscillating energy of every MS mode in a ferrite-disk sample be represented as the energy of an ideal gas of magnons? A priori, one can expect a negative answer to the last question since, as we discussed above, the energy variation of MS modes is determined not only by the magnetic field but also by the electric field induced in accordance with the Faraday law. More detailed proof will be done below.

Quantization of magnetization may take place if there is a spectral problem for a magnetization field or if a magnetization field can be expressed by the compete-set scalar functions. Both these cases do not occur for MS oscillations. For the biasing magnetic field directed along the z-axis, the rf magnetization in the MS description is written in cylindrical coordinates as [6]:

$$m_\rho = -\chi \frac{\partial \psi}{\partial \rho} - i\chi_\alpha \frac{1}{\rho} \frac{\partial \psi}{\partial \alpha},$$
$$m_\alpha = i\chi_\alpha \frac{\partial \psi}{\partial \rho} - \chi \frac{1}{\rho} \frac{\partial \psi}{\partial \alpha}, \qquad (37)$$
$$m_z = 0,$$

where $\chi$ and $\chi_a$ are, respectively, diagonal and off-diagonal components of tensor $\tilde{\chi}$. For a dominant "thickness" mode described by the known function $\tilde{\xi}(z)$ and $q$ "flat" mode described by $\tilde{\varphi}_q(\rho,\alpha)$ [see Eqn. (19)], one has

$$(m_\rho)_q = A_q \tilde{\xi}\left(-\chi \frac{\partial \tilde{\varphi}_q}{\partial \rho} - i\chi_\alpha \frac{1}{\rho} \frac{\partial \tilde{\varphi}_q}{\partial \alpha}\right),$$
$$(m_\alpha)_q = A_q \tilde{\xi}\left(i\chi_\alpha \frac{\partial \tilde{\varphi}_q}{\partial \rho} - \chi \frac{1}{\rho} \frac{\partial \tilde{\varphi}_q}{\partial \alpha}\right), \qquad (38)$$
$$(m_z)_q = 0.$$

Based on these expressions, one can introduce the corresponding "in-plane" differential operators:

$$\hat{m}_\rho = -\chi \frac{\partial}{\partial \rho} - i\chi_\alpha \frac{1}{\rho} \frac{\partial}{\partial \alpha},$$
$$\hat{m}_\alpha = i\chi_\alpha \frac{\partial}{\partial \rho} - \chi \frac{1}{\rho} \frac{\partial}{\partial \alpha}, \qquad (39)$$

Now we define the operator:

$$\hat{m}_\perp^2 = \hat{m}_\rho^2 + \hat{m}_\alpha^2 \qquad (40)$$

This operator is expressed as:

$$\hat{m}_\perp^2 = (\chi^2 - \chi_a^2)\left(\frac{\partial^2}{\partial \rho^2} + \frac{1}{\rho^2}\frac{\partial^2}{\partial \alpha^2}\right) = \gamma M_0 \left(\nabla_\perp^2 - \frac{1}{\rho}\frac{\partial}{\partial \rho}\right) \qquad (41)$$

Let us suppose, as a particular case, that a lateral cylindrical surface of a ferrite disk is a perfect magnetic wall. In this case one has the same domains of definition of operators $\hat{m}_\perp^2$ and $\nabla_\perp^2$. Unlike energy operator $\hat{F}_\perp$, operator $\hat{m}_\perp^2$ is not proportional to the Laplace "in-plane" operator $\nabla_\perp^2$. Evidently, operators $\hat{F}_\perp$ and $\hat{m}_\perp^2$ *do not commute with each other*. It means that for every oscillating MS "flat" mode $q$ (with dominant "thickness" mode $p$), energy $E_q$ described by the MS-potential function [see Eqn. (31)] is not equal to the oscillating energy of the rf magnetization in a ferrite sample. It means also that for every energy eigenstate $E_q$, the value $m_q^2 = \left(m_\rho\right)_q^2 + \left(m_\alpha\right)_q^2$ is not a constant (stable) quantity. So the *energy eigenstates $E_q$ are not characterized by a stable amount of magnons*. One can suppose that in accordance with Eqn. (38), for every (normalized) MS-potential mode there is a "magnetization mode". But the "magnetization modes" are not orthonormalized ones. In other words, energy of magnetization does not belong to the orthonormal space of energy. The fact that in our case we have quantization of MS energy (energy eigenstates of MS oscillations) does not stipulate quantization of the magnetization energy in MS-wave oscillations.

The above statement that one does not have quantization of the magnetization energy in MS-wave oscillations can also be illustrated in another way. Let us suppose that operator $\hat{m}_\perp^2$ has eigenvalues, which we denote as $m^2$, and "flat" scalar eigenfunctions $\tilde{u}$ different from function $\tilde{\varphi}$. In this case, we have the equation:

$$\hat{m}_\perp^2 \tilde{u} = m^2 \tilde{u} \tag{42}$$

or

$$\frac{\partial^2 \tilde{u}}{\partial \rho^2} + \frac{1}{\rho^2}\frac{\partial^2 \tilde{u}}{\partial \alpha^2} - \frac{m^2}{\chi^2 - \chi_a^2}\tilde{u} = 0 \tag{43}$$

Here $m^2 = m_\rho^2 + m_\alpha^2$ is the quantity defining density of magnetization energy [6]. Because a cylindrical symmetry and in supposition that functions $\tilde{u}$ and $\tilde{\varphi}$ might be at the same azimuth symmetry, we have $\tilde{u} \sim e^{-iv\alpha}$. Eqn. (43) is written as:

$$\frac{\partial^2 \tilde{u}}{\partial \rho^2} - \frac{v^2}{\rho^2}\tilde{u} + Q^2 \tilde{u} = 0 \tag{44}$$

where we denoted:

$$Q^2 = \frac{m^2}{\chi_a^2 - \chi^2} \tag{45}$$

Now let us introduce a new real quantity $n$, which is defined from the following equation:

$$\frac{4n^2 - 1}{4} = v^2 \tag{46}$$

Based on Eqn. (46) we rewrite Eqn. (44) as:

$$\frac{\partial^2 \tilde{u}}{\partial \rho^2} + \left(Q^2 - \frac{4n^2 - 1}{4\rho^2}\right)\tilde{u} = 0 \tag{47}$$

One has the following solution for function $\tilde{u}$ [15]:

$$\tilde{u} = \text{const}\, \sqrt{\rho}\, J_n(Q\rho) e^{-iv\alpha} = \text{const}\, \sqrt{\rho}\, J_{\pm\frac{1}{2}\sqrt{4v^2+1}}(Q\rho) e^{-iv\alpha} \tag{48}$$

For the essential boundary conditions, functions $\tilde{\varphi}$, being expressed by Bessel's functions, are mutually orthogonal [7,16]. Functions $\tilde{u}$ are not expressed by Bessel's functions. There is no formal evidence of orthogonality of functions $\tilde{u}$. Moreover, if one suppose that functions $\tilde{\varphi}$, form a complete set of scalar functions and that operators $\hat{m}_\perp^2$ and $\nabla_\perp^2$ have the same domains of definition, another-type scalar functions $\tilde{u}$ cannot be the complete-set functions.

## 6. MS MODES AS AN IDEAL GAS OF "LIGHT MAGNONS"

The above analysis shows that there is no quantization of the magnetization energy in MS-wave oscillations. At the same time, every MS-wave oscillating mode is characterized by a certain energy level. This energy is not the energy of an ideal gas of magnons. One can suppose that in our case of MS oscillations in a ferrite disk resonator, there exist certain quasiparticles having quantization of energy and characterizing by certain *effective masses*. We, conventionally, will name these quantum quasiparticles as the "*light magnons*" (lm). The meaning of this term arises from the fact that an effective mass of the "light magnons" should be much less than an effective mass of the ("real", "heavy") magnons – the quasiparticles existing due to the exchange interaction. In our description of MS oscillations we neglect the exchange interaction and the "magnetic stiffness" is characterized by the "weak" dipole-dipole interaction [4-6]. The states of the "light magnons" are described based on the so-called *transitional eigenfunctions* [17]. For these transitional eigenfunctions energy is proportional to a squared wavenumber [17]. In our case this is a squared wavenumber of a propagating MS mode. For MS waveguide mode $n$, the number of "light magnons" in a MS waveguide section is defined from Eqns. (12) and (29) as the ratio: $\frac{E_n}{\varepsilon}$. When we juxtapose Eqn. (22) with the Schrödinger equation for "free particles" ($a^{(2)} \equiv 0$), we get the following expression for an effective mass of a "light magnon":

$$m_{eff}^{(lm)} = \frac{i\hbar}{2a^{(1)}} \qquad (49)$$

For a monochromatic wave process ($\psi \sim e^{i\omega t}$) we obtain for mode $n$:

$$\left(m_{eff}^{(lm)}\right)_n = \frac{\hbar}{2}\frac{\beta_n^2}{\omega} \qquad (50)$$

This expression looks very similar to an effective mass of the ("real", "heavy") magnon for spin waves with the quadratic character of dispersion [6].

The feature of an infinite-ferrite-rod MS-wave waveguide is the fact that in this waveguide structure there are two cutoff frequencies $\omega_1, \omega_2$. For a given frequency in the frequency region between the cutoff frequencies ($\omega_1 \leq \omega \leq \omega_2$), there is a complete discrete spectrum of propagating MS modes. For given frequency $\omega'$ ($\omega_1 \leq \omega' \leq \omega_2$), one has a flow of quasiparticles with different "effective masses" and different "kinetic energies". For another frequency $\omega'' \neq \omega'$ ($\omega_1 \leq \omega'' \leq \omega_2$) we have a flow of another quasi-particles differing from previous ones by "effective masses" and "kinetic energies". At a certain frequency, the total energy of non-interacting quasiparticles is equal to a sum of energies of separate quasiparticles:

$$E_{tot}^{(lm)} = \sum_n E_n^{(lm)} \tag{51}$$

In a case of a ferrite disk resonator, there is a certain quantity of the propagation constant for every resonance frequency. For MS "flat" mode $q$ (and the dominant "thickness" mode) oscillating at frequency $\omega_q$, one has:

$$\left(m_{eff}^{(lm)}\right)_q = \frac{\hbar}{2} \frac{\left(\beta_q^{(F)}\right)^2}{\omega_q} \tag{52}$$

One can classify the types of "light magnons" with respect to the numbers $\nu$ characterizing the azimuth distribution. For $\nu = 1$ we have the dipole-type "light magnon". For $\nu = 2$ we have the quadrupole-type "light magnon". For higher integer quantities $\nu$ we have the multipole-type "light magnons". For every concrete type of a "light magnon", the energy level is determined by the radial mode index $q$ of the $\widetilde{\varphi}$-function.

Based on the above consideration, we can discuss the physical model of the observed multiresonance MS oscillations in a flat ferrite disk resonator. For a given type of a "light magnon" (characterizing by number $\nu$), we have eigen space oscillations (with respect to z-coordinate) of the MS-potential function in a "rectangular potential well". This well is due to a negative diagonal component $\mu$ of a permeability tensor inside a ferrite. For different frequencies or different bias magnetic fields (in regions, where $\mu < 0$), there are different properties of the continuum-model "material" (characterizing by parameter $\mu$), "filling" the potential well. As a result, one has a set of energy eigenstates of MS oscillations for every type of a "light magnon". Multi-resonance MS oscillations in a disk ferrite resonator are due to motion of "light magnons" between two planes: $z = 0$ and $z = h$. Velocity of this "light magnon" motion is not a phase velocity of MS waves along z-axis. Similar to a free electron, when the particle velocity is not a phase velocity of de Brogilie's wave, but the velocity of the wave packet, in our case we should consider the group velocity. This fact was taken into account in our analysis: the normalized energy of MS oscillations [see Eqn. (31)] was derived based on consideration of the *quasimonochromatic* MS-wave process [7].

The wave packet describes the localization of the particle. The position of the "light magnon" is characterized by the probability density distribution function $\psi\psi^*$ [7]. So the "light magnons" are quantum quasiparticles having lack of localization in space of a ferrite sample. At the same time, a ferrite resonator (containing "light-magnons") is a localized particle with respect to the free-space-electromagnetic-wavelength scales.

## 7. CALCULATIONS

Based on the above model of the "light magnon" oscillations in a ferrite disk resonator, we made a series of calculations. The graphical solutions of Eqns. (33) and (34) with respect to the DC magnetic field, are illustrated in Fig. 1. The solutions were obtained for the main "thickness mode" and different "in-plane (flat) modes" calculated for Bessel functions of orders $\nu = 1, 2, 3$ and with a number of radial variations ($q$ numbers). An analysis was made with use of the disk data given in paper [3]:

$4\pi M_s = 1790\,G$, $2R = 3.98\,mm$, $h = 0.284\,mm$. The working frequency is $\frac{\omega}{2\pi} = 9.51\,GHz$. The feature of Eqn. (34) is the fact that in the point where $\mu = -1$, this equation becomes an identity. So at the DC magnetic field corresponding to the quantity $\mu = -1$ one has a break.

For known quantities $\beta_q^{(F)}$, the energy levels were calculated based on Eqn. (31). Figs. 2 demonstrate the positions of quantities $E_q^{(lm)}$ – the normalized energies of the "light-magnon" collection – corresponding to different "flat modes". The modes are characterized by numbers $q$ and quantities $\nu$. In accordance with the above classification, Fig. 2 (a) shows the energy levels for the dipole-type "light magnon" collection ($\nu = 1$), Fig. 2 (b) shows the energy levels for the quadrupole-type "light magnon" collection ($\nu = 2$), and Fig. 2 (c) – for the hexapole-type "light magnon" collection ($\nu = 3$). Effective masses of "light magnons" corresponding to different MS modes were calculated based on Eqn. (52). The results are shown in Figs. 3 (a), (b), (c) for different-type "light magnons". The behaviors of "light magnons" in a ferrite disk resonator can be illustrated by MS-potential distributions with respect to axial, radial and azimuth coordinates. For some of modes, these distributions are shown, respectively, in Figs. 4, 5, and 6.

In MS-wave processes, the MS-potential function can be considered as the probability distribution function [7]. The probability density distribution function $\psi\psi^*$ shows the "light magnon" distribution. In particular, the "light magnon" distribution in a plane of a ferrite disk resonator is defined by the probability density distribution function $\widetilde{\varphi}\widetilde{\varphi}^*$. Fig. 7 shows the probability density distribution functions $\widetilde{\varphi}\widetilde{\varphi}^*$ for different dipole-type ($\nu = 1$) "light magnons". For other-type "light magnons" $\widetilde{\varphi}\widetilde{\varphi}^*$ at the same quantities $q$, the probability density distribution functions are similar to those shown in Fig. 7.

The main feature of the magnetic-field spectrum shown in Fig. 1 is the fact that high-order peaks correspond to lower quantities of the DC magnetic field. Physically, the situation looks as follows. Let $H_0^{(1)}$ and $H_0^{(2)}$ be, respectively, the upper and lower values of a bias magnetic field corresponding to the borders of a region, where $\mu < 0$ [6]. Suppose we have a bias field $H_0^{(1)}$. When we put a ferrite sample into this field, we supply it with the energy: $4\pi M_0 H_0^{(1)}$. To some extent, this is a pumping-up energy. *Starting from this point,* we can excite the entire spectrum: from the main mode to the high-order modes. As we move from value $H_0^{(1)}$ to value $H_0^{(2)}$, the energy surplus goes over to the high-order-mode excitation.

Let us calculate the total depth of a "potential well". For the working frequency $\frac{\omega}{2\pi} = 9.51\, GHz$ and saturation magnetization $4\pi M_0 = 1792\, Gauss$ – the data of the Yukawa and Abe's experiments [3] – the depth is calculated as:

$$\Delta U = 4\pi M_0 \left( H_0^{(1)} - H_0^{(2)} \right) = 780\, Oersted \times 1792\, Gauss = 1.4 \cdot 10^6 \frac{ergs}{cm^3} = 14 \cdot 10^4 \frac{Joule}{m^3} \qquad (53)$$

As a value of a bias magnetic field decreases, the "particle" obtains the *higher levels of negative energy*. For example, the first three levels ($q = 1, 2, 3$) of negative potential energy for dipole-type ($\nu = 1$) "light magnons" calculated as:

$$U_q = -4\pi M_0 \left( H_0 \big|_{q=1,2,3} - H_0^{(1)} \right), \qquad (54)$$

are equal, respectively, to $-1.5 \times 10^4 \frac{Joule}{m^3}$, $-4.3 \times 10^4 \frac{Joule}{m^3}$, $-6.2 \times 10^4 \frac{Joule}{m^3}$ (For these calculations the quantities $H_0 \big|_{q=1,2,3}$ were taken from Fig. 1 as the first-three-peak positions in the magnetic-field spectra for $\nu = 1$).

The situation is very resembling the increasing a negative energy of the hole in semiconductors when it "moves" from the top of a valence band [18]. In classical theory, negative-energy solutions are rejected because they cannot be reached by a continuous loss of energy. But in quantum theory, a system can jump from one energy level to a discretely lower one; so the negative-energy solutions cannot be rejected, out of hand. When one continuously varies the quantity of the DC field $H_0$, for a given quantity of $\omega$, one sees a discrete set of absorption peaks. It means that one has the discrete-set levels of potential energy. This is a very crucial fact that the jumps between the potential levels are controlled (are governed) by the discrete transitions between the quantum states of the "light magnons".

The energies found from Eqn. (31) can be considered as "kinetic energies". At the same time, the fact that the spectral properties are exhibited with respect to quantities of a bias magnetic field means variations of "potential energy" of a ferrite sample. So the quantized quantities of the "light magnon" kinetic energy take place only on certain discrete levels of potential energy. Based on the above analysis, we can find certain appropriateness between the "light-magnon" energy levels and the potential energy levels. Such appropriateness between the $E_q^{(lm)}$ and $|U_q|$ levels is illustrated in Figs. 8 for different-type "light magnons".

The number of "light magnons" available on every potential level is defined as a ratio of Eqns. (31) and (16). We have:

$$N_q^{(lm)} = \frac{E_q^{(lm)}}{\varepsilon_q} = \frac{g\mu_0 \left(\beta_q^{(F)}\right)^2}{2\hbar\omega_q} = \frac{g\mu_0}{\hbar^2}\left(m_{eff}^{(lm)}\right)_q \qquad (55)$$

So the number of "light magnons" on the energy level is proportional to the quantity of an effective mass corresponding to this level. This should not be considered as a surprising situation since for a constant operating frequency and varying bias magnetic field the energy of the individual "light magnon" $\varepsilon_q$ is the same.

# 8. DISCUSSION AND CONCLUSION

Ferromagnetic resonators with MS oscillations can be considered in microwaves as point (with respect to the external electromagnetic fields) particles. MS oscillations in a small ferrite disk resonator can be characterized by a *discrete spectrum of energy levels*. This fact allows analyzing the MS oscillations similarly to quantum mechanical problems. A special interest of energy spectra of such a small structural element – artificial magnetic atoms – may be found in the fields of microwave artificial composite materials, microwave spectroscopy, and, probably, quantum computation.

From the above analysis it follows that the energy quantization (described by the MS potential properties) can be regarded as a collective effect of quasiparticles – the "light magnons". In other words, the MS-wave phenomena in a special *macrodomain* – a ferrite disk resonator – can be simply reduced to the case of a many-particle correlated system. Our estimations give that the "light magnon" effective mass $m_{eff}^{(lm)}$ (for YIG disk resonators with parameters corresponding to the data of experiments [3]) is a very small quantity, which is about $10^8$ times less than the free electron mass. At the same time, the ("heavy") magnon mass in YIG is approximately 6 times more than the free electron mass [6]. This fact is very clear since MS oscillations take up an "intermediate position" between the electromagnetic and exchange oscillations. From one hand, to describe these oscillations we can put the phenomenological exchange constant equal to zero [in this case the ("heavy") magnon mass becomes equal to infinity]. From the other hand, the MS oscillations are described in neglect of the electric displacement current in the Maxwell equations. So the quantity of the "light-magnon" effective mass should be between the ("heavy") magnon effective mass (very big) and zero mass of photon.

Discussing the above model and the obtained calculation results, we have to point out that in a series of his previous works, Morgenthaler had demonstrated already that one can introduce the notion of quasiparicles and use the time-independent Schrödinger equation for certain cases of MS modes [19,20].

We have to give now a clear explanation what is a strong difference between our model and Morgenthaler's ideas and why based just only on our consideration of confinement phenomena one can obtain the magnetic artificial atoms.

As we have shown in this paper, our analysis gives us the possibility to get the orthonormality relation for MS-potential wave functions, to formulate the energy eigenvalue problem, to calculate the energy spectra for a whole ferrite-particle system. So our consideration is based on a real physical picture. The Morgenthaler model also provides the reader with the method of the quantum-mechanical analog for a certain wave function. This wave function is correlated with the rf magnetization. Contrary to our analysis, however, one cannot see that the used wave functions constitute the orthonormal functional basis with the norms corresponding to energies. Thus this implies just only the method useful in developing a certain qualitative understanding of mode properties, but not in developing really new physical systems. In Section 5 of our paper we showed that for MS modes one does not have neither real quantization of magnetization, nor the possibility to express the magnetization field by the complete-set scalar wave functions.

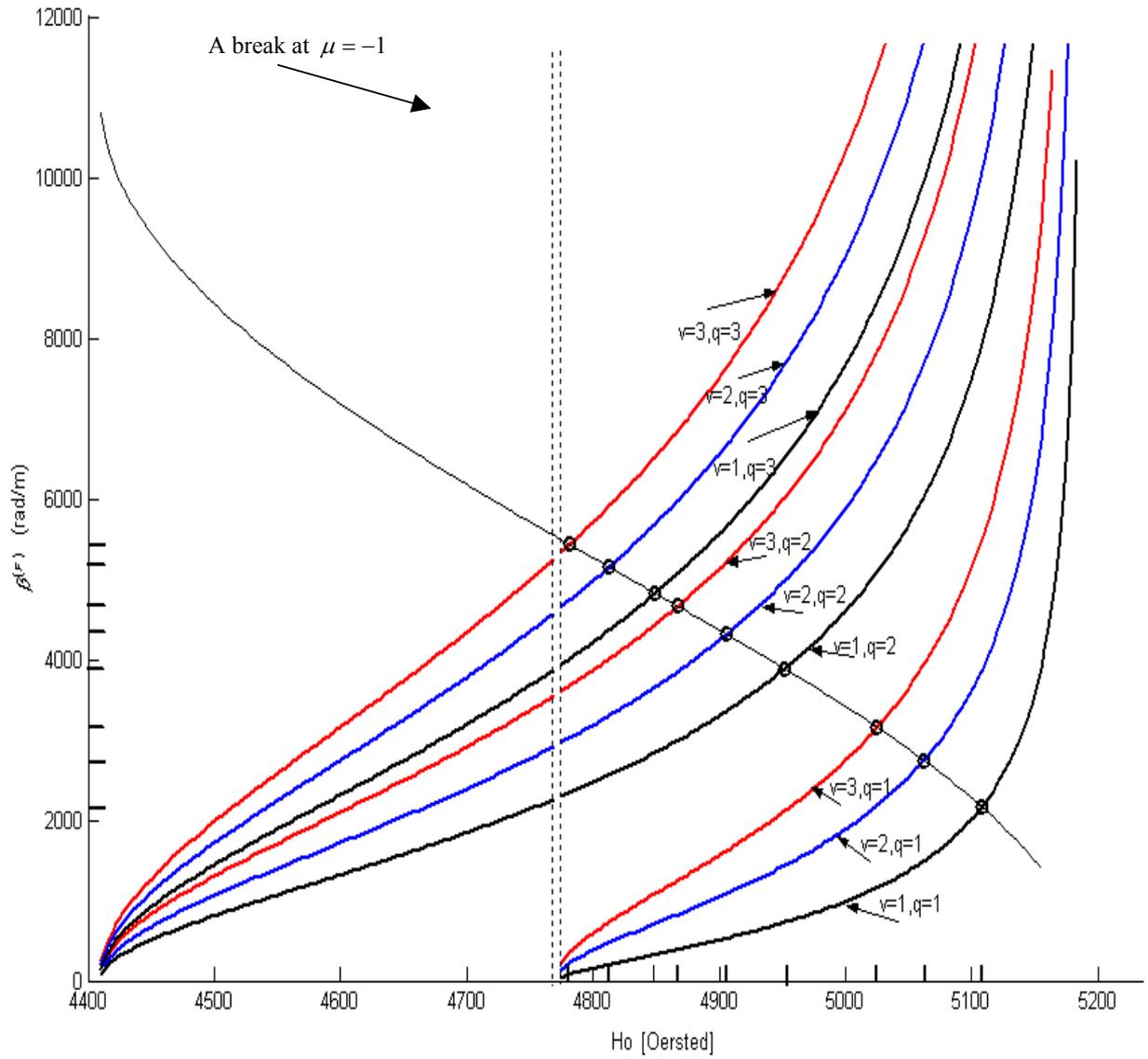

Fig.1. The graphical solutions of Eqns. (33) and (34) with respect to the DC magnetic field for the main "thickness mode" and different "in-plane (flat) modes".

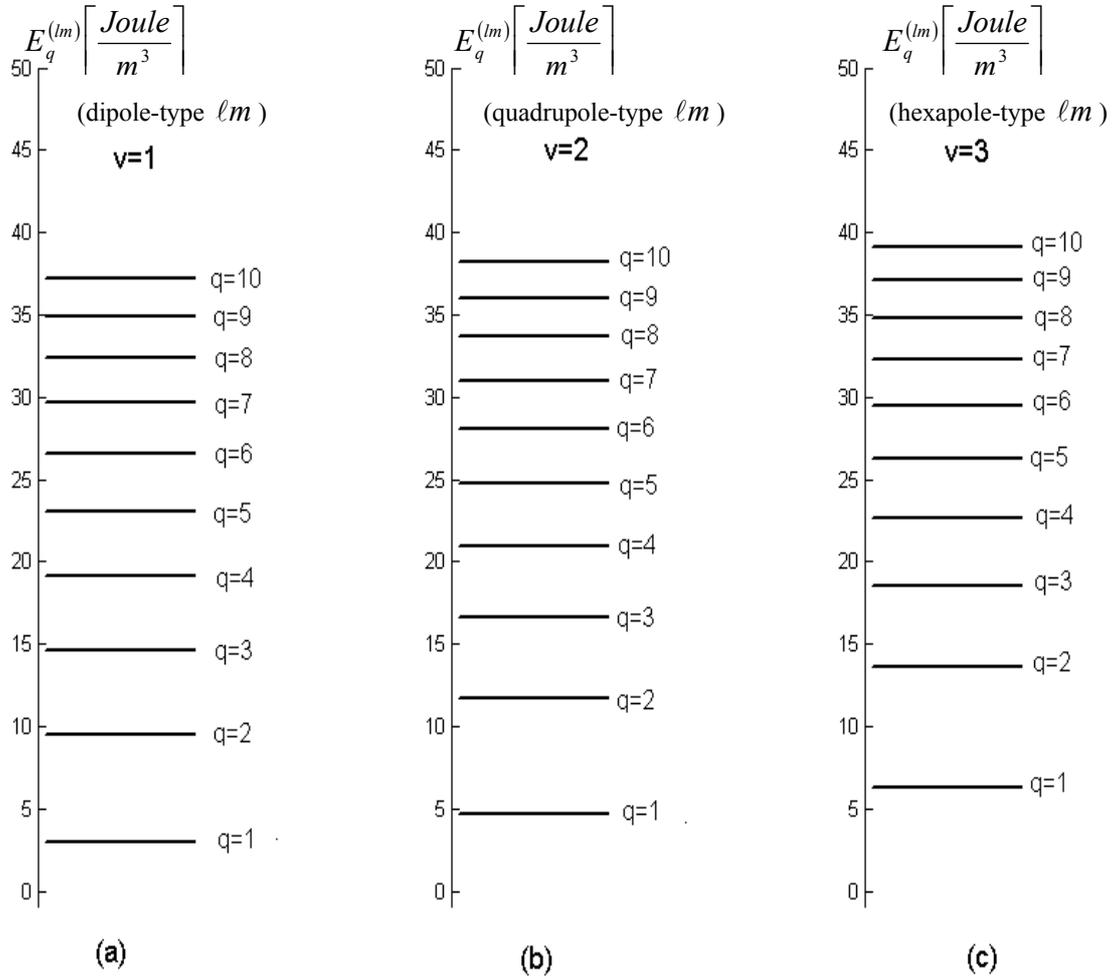

Fig. 2. The energy levels for different-type ″light magnons″ collections
(a) the energy levels for the dipole-type ″light magnons″ collection ($\nu = 1$),
(b) the energy levels for the quadrupole-type ″light magnons″ collection ($\nu = 2$),
(c) the energy levels for the hexapole-type ″light magnons″ collection ($\nu = 3$).

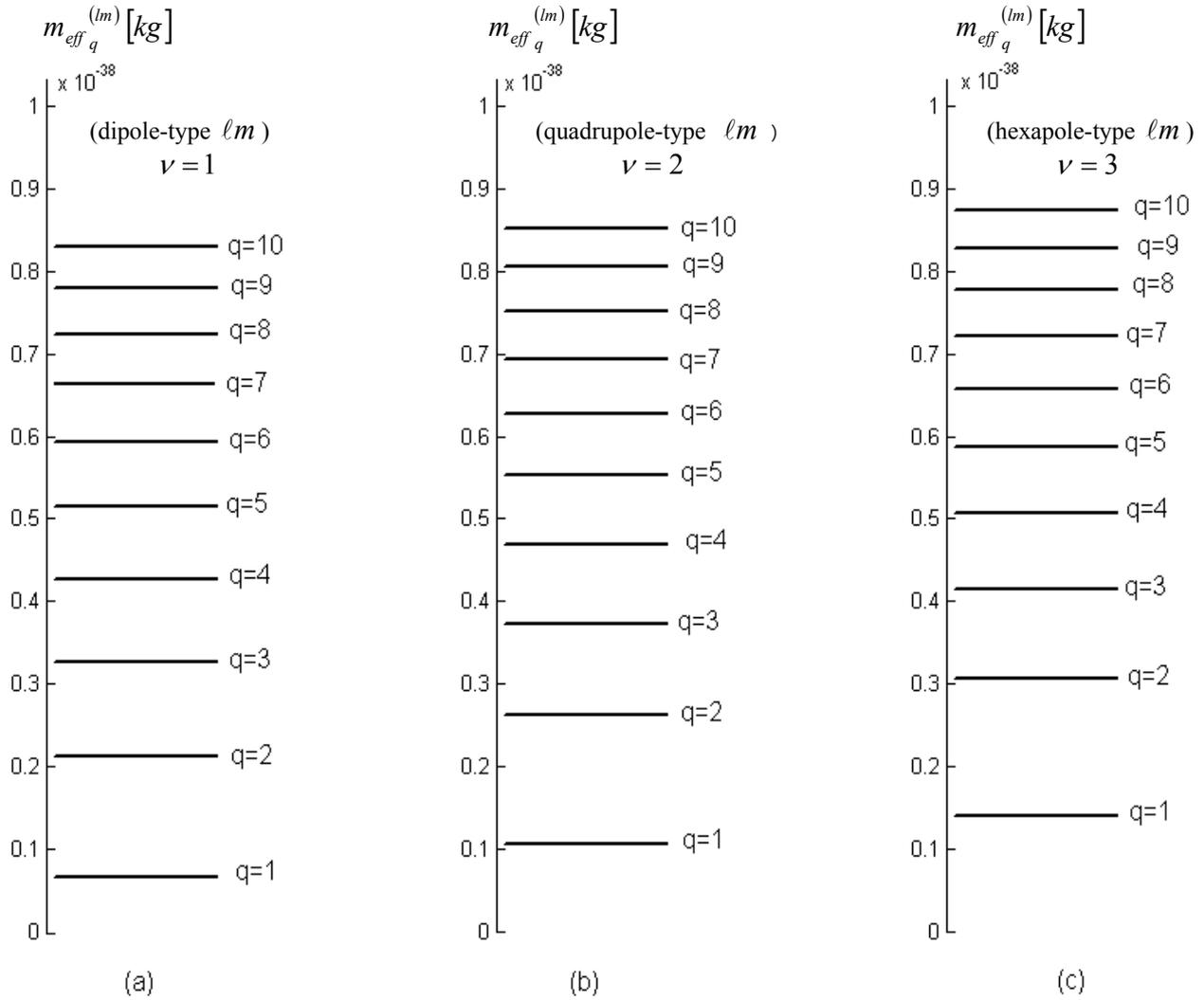

Fig.3. Effective masses of ″light magnons″.
  a) the dipole-type ″light magnons″ ($\nu = 1$),
  b) the quadrupole-type ″light magnons″ ($\nu = 2$),
  c) the hexapole-type ″light magnons″ ($\nu = 3$).

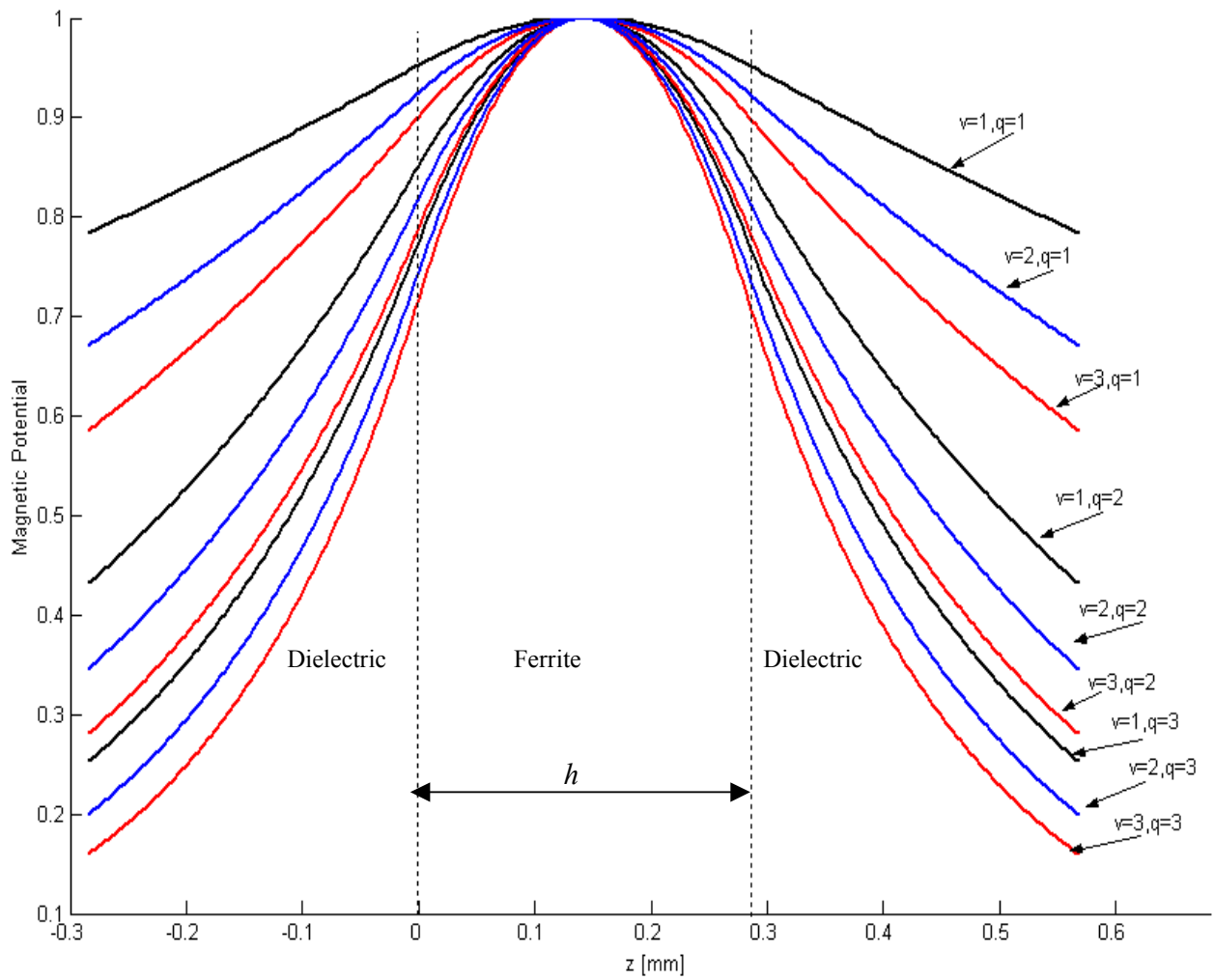

Fig. 4. MS-potential distribution with respect to an axial coordinate

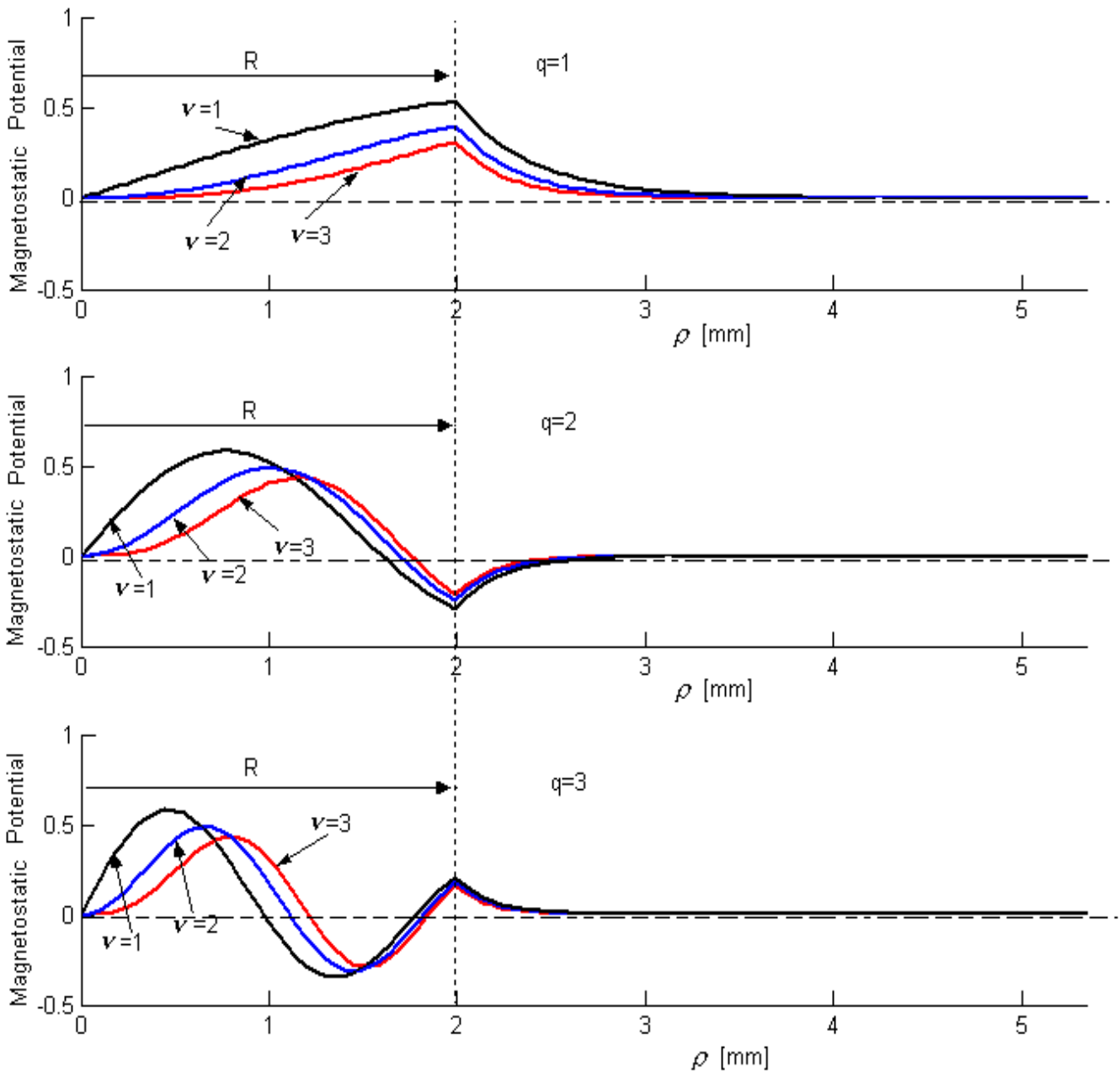

Fig. 5. MS-potential distribution with respect to a radial coordinate

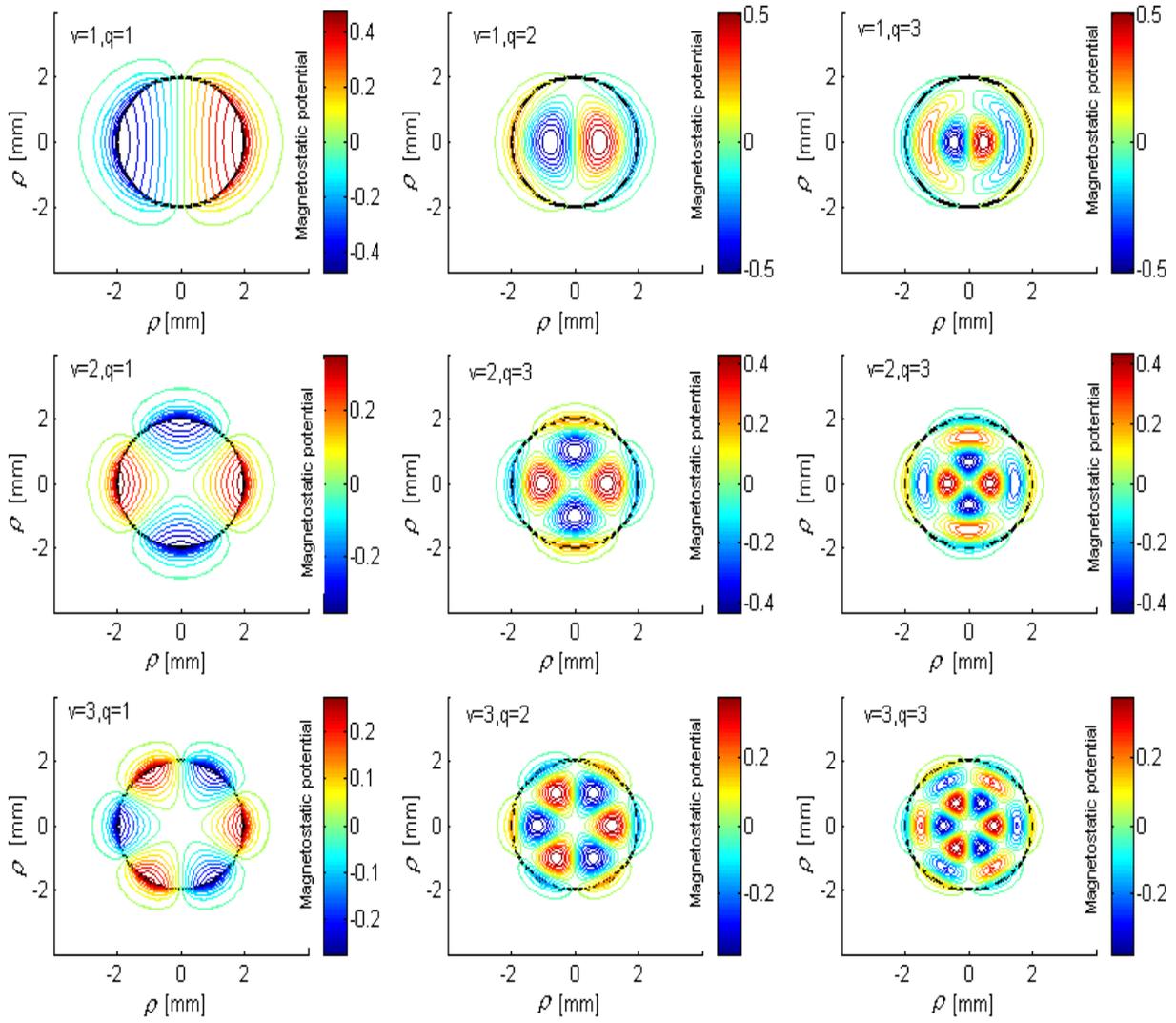

Fig.6. MS-potential distribution with respect to an azimuth coordinate

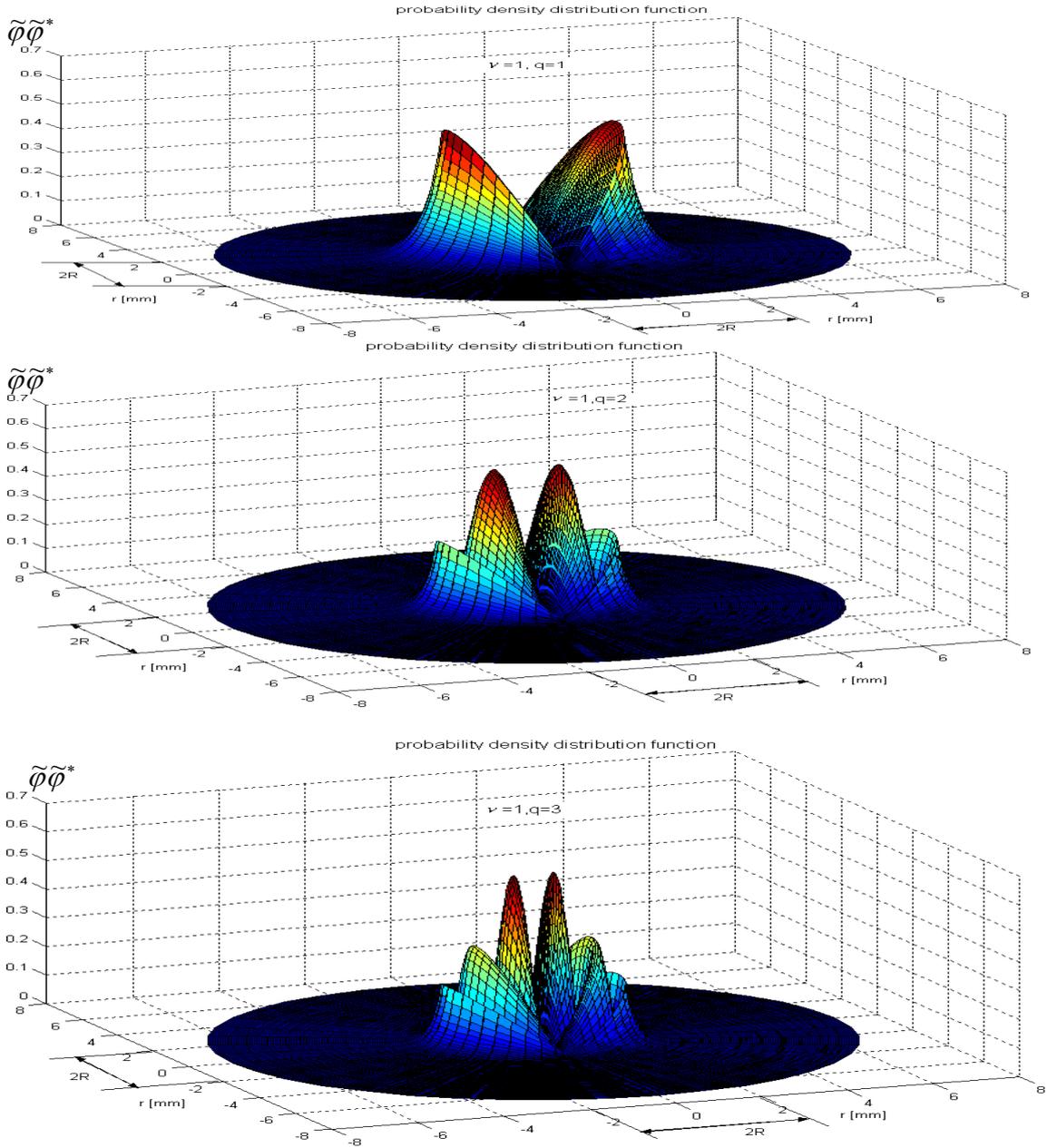

Fig.7. Probability density function $\widetilde{\varphi}\widetilde{\varphi}^*$ for dipole-type ($\nu = 1$) ″light magnons″

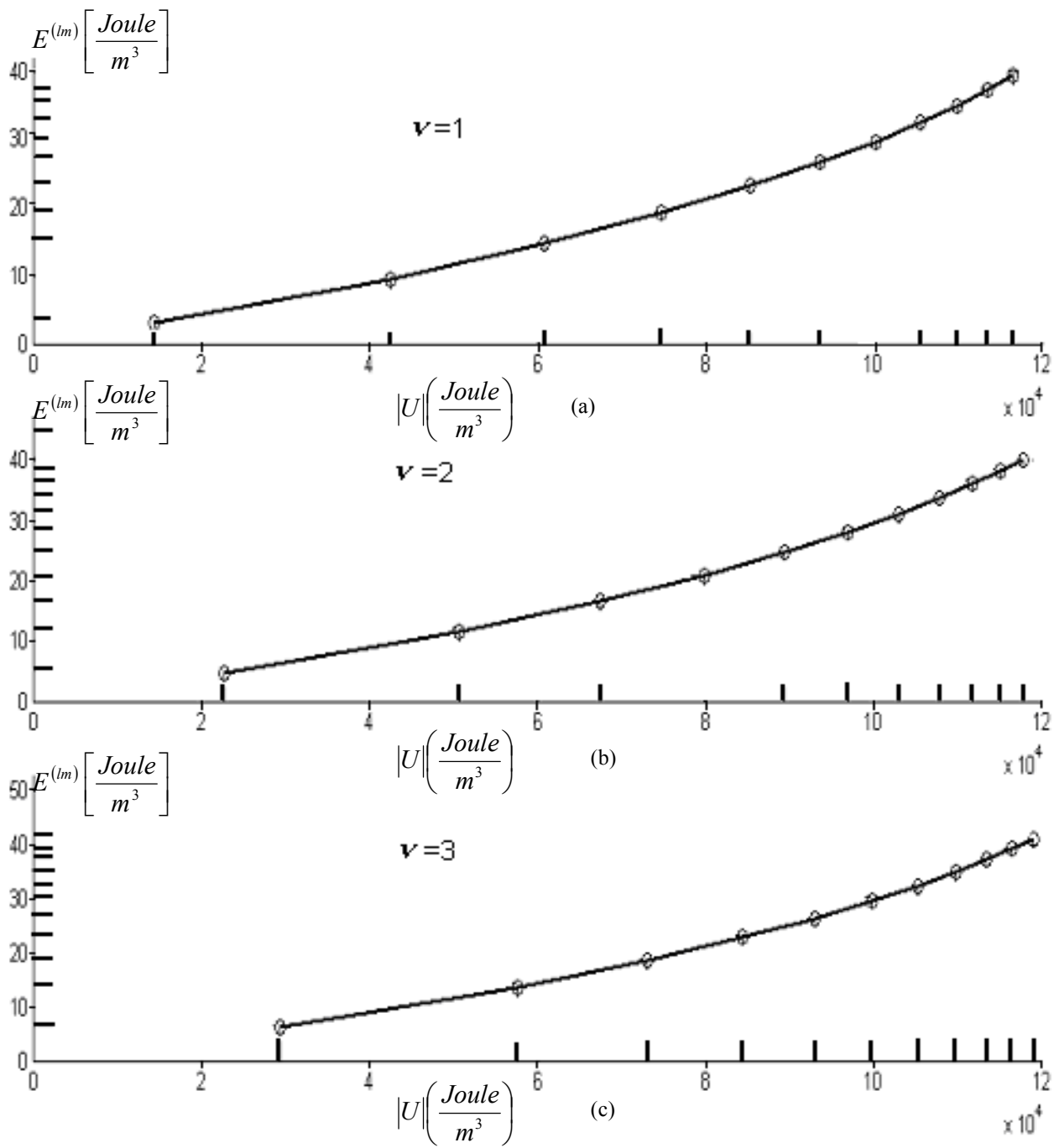

Fig. 8. Appropriateness between the $E_q^{(lm)}$ and $|U_q|$ levels
  a) for dipole-type "light-magnons" $(\nu = 1)$,
  b) for quadrupole-type "light-magnons" $(\nu = 2)$,
  c) for hexapole-type "light-magnons" $(\nu = 3)$